\documentclass[usenatbib,usegraphicx]{mn2e}

\title[Asymmetric Double Lens J0316+4328]{J0316+4328: a Probable ``Asymmetric Double'' Lens}
\author[E. R. Boyce et al.]{E.~R.~Boyce$^{1}$\thanks{E-mail: edward.boyce@manchester.ac.uk (ERB)}, S.~T.~Myers$^{2}$, I.~W.~A.~Browne$^{1}$, W.~J.~Stroman$^{2,3}$ and N.~J.~Jackson$^{1}$\\
$^{1}$University of Manchester, Jodrell Bank Observatory, Macclesfield, Cheshire SK11 9DL\\
$^{2}$National Radio Astronomy Observatory, PO Box O, Socorro, NM 87801, United States\\
$^{3}$Iowa State University, Department of Physics and Astronomy, 12 Physics Hall, Ames, IA 50011, United States}
\begin{document}

\date{Accepted 2007 July 16. Received 2007 June 30; in original form 2007 June 04}

\pagerange{\pageref{firstpage}--\pageref{lastpage}} \pubyear{2007}

\maketitle

\label{firstpage}

\begin{abstract}
We report a probable gravitational lens J0316+4328, one of 19 candidate
asymmetric double lenses (2 images at a high flux density ratio) from CLASS. 
Observations with the Very Large Array (VLA), MERLIN and the Very Long Baseline
Array (VLBA) imply that J0316+4328 is a lens with high confidence.
It has 2 images separated by 0\farcs40, with 6~GHz flux densities of
62~mJy and 3.2~mJy. The flux density ratio of $\sim19$ (constant over the 
frequency range 6-22~GHz) is the largest for any 2 image gravitational lens. 
High resolution optical imaging and deeper VLBI maps should confirm the 
lensing interpretation and provide inputs to detailed lens models. The unique 
configuration will give strong constraints on the lens galaxy's mass profile.
\end{abstract}


\section[]{Introduction}

Galaxy-scale gravitational lenses are valuable tools for the study of galactic 
structure. The locations and brightnesses of the images provide constraints on 
the overall lens galaxy profile, while remaining anomalies and time variations 
in the brightnesses due to micro-lensing probe substructure on scales from 
dark matter clumps \citep{mao98a, dal02a} and satellite galaxies 
\citep{mck07a} to stars \citep{sch02a}. 
Lens images are sensitive to the surface densities at the locations where they 
form, generally at galactocentric radii of a few kpc for bright lens images.
Measuring the overall lens galaxy profile over a range of scales requires more 
information, which can be provided by stellar velocity dispersions at radii 
$<1$~kpc \citep{koo06a} or weak lensing at radii of tens of kpc 
\citep{gav07a}. Lenses with central images measure the density profile on many
scales using strong lensing alone, giving strong constraints at radii of 
$\sim100$~pc, including any contribution from a central super-massive 
black hole \citep{win04a}. 

For lensed quasars, the best measurement of the lens galaxy profile can
be done if the quasar is radio-loud. The size of the emitting region is larger
in the radio than in optical or X-ray \citep{koo00a}, making the image
brightnesses less susceptible to microlensing, while central images are 
strongly affected by dust absorption and confusion in the optical regime
but not in the radio frequency range 5-40~GHz \citep{win04a}.
The largest sample of radio-loud lenses has been provided by CLASS
\citep{mye03a,bro03a}, which found 22 instances of a bright radio source
lensed by a foreground galaxy.

A crucial step in CLASS was the rejection of double sources with a
flux density ratio $>10:1$ \citep{mye03a} after the initial 8.4~GHz
VLA snapshot. This was done for two reasons; for statistical
completeness of the lens sample it was necessary to be sure that all
secondary components could be detected reliably and it also
significantly reduced the amount of follow-up observations required. A
secondary compact component of similar brightness to the primary is
likely to be a lensed image, while a secondary component much fainter
than the primary is most likely to be weak structure belonging to the
primary source. This strategy necessarily excluded ``asymmetric
doubles'': 2 image lenses with a high flux density ratio. Such
lenses are of interest because they are most likely to host observable
central images \citep{mao01a, bow04a}, while even the absence of a central 
image gives the strongest constraints in an asymmetric double
\citep{rus01a, boy06b, zha07a}.  Though the recognition of weak
secondaries with ratios $>10:1$ in CLASS may not have been
100\% reliable, many are detectable. We have, therefore, conducted a
program to identify such lenses by following up promising candidates
which were initially rejected only on the basis of a high flux density
ratio.

\section[]{Survey and Observations}

\begin{table*}
 \centering
  \caption{Summary of VLA observations. 
    Flatter and steeper refer to the spectral indices of weaker components 
    over the observed frequencies, relative to the primary compact source. 
    e.g. J0008+6837 had 2 compact radio components, the fainter of these 
    had a steeper radio spectrum between 5 and 8.5~GHz. Most of
    our candidates were typical radio sources with a flat-spectrum, compact 
    core and a steeper spectrum, often extended, lobe or knot in a jet.
    Only J0316+3350, J0935+0719 and J0958+2948 could be considered lens
    candidates from the VLA observations. 
  \label{vla}}
  \begin{tabular}{@{}lrrrrrl@{}}
    \hline
    Source & R.A. & Dec. & \multicolumn{3}{c}{Observing Time (mins.)} & Notes\\
    & \multicolumn{2}{c}{(J2000)} & 5 GHz & 8.5 GHz & 22 GHz & \\
    \hline
    J0008+6837 & 00 08 33.5 & 68 37 22 & 14 & 14 & - & Rejected: 2nd component steeper\\
    J0152+3350 & 01 52 34.6 & 33 50 33 & 6 & 8 & - & Rejected: 2nd component steeper\\
    J0316+3350 & 03 16 50.9 & 43 28 19 & 11 & 12 & 17 & Possible Lens from VLA observations\\
    J0644+5955 & 06 44 04.7 & 59 55 21 & 6 & 7 & - & Rejected: 2nd and 3rd components extended, steeper\\
    J0812+4041 & 08 12 03.0 & 40 41 08 & 18 & 18 & - & Rejected: 2nd and 3rd components extended, steeper\\
    J0852+5922 & 08 52 30.0 & 59 22 50 & 18 & 19 & - & Rejected: 2nd component extended, steeper\\
    J0903+4651 & 09 03 04.0 & 46 51 04 & 4 & 5 & 45 & Rejected: 2nd component extended\\
    J0935+0719 & 09 35 01.1 & 07 19 19 & 12 & 14 & 23 & Possible Lens from VLA observations\\
    J0938+3934 & 09 38 39.2 & 39 34 20 & 18 & 18 & - & Rejected: 2nd component extended, steeper\\
    J0958+2948 & 09 58 58.9 & 29 48 04 & 12 & 13 & 21 & Possible Lens from VLA observations\\
    J1131+5146 & 11 31 16.5 & 51 46 34 & 18 & 19 & - & Rejected: 2nd component extended, steeper\\
    J1252+1910 & 12 52 27.8 & 19 10 38 & 18 & 19 & - & Rejected: 2nd component extended, steeper\\
    J1253+6304 & 12 53 17.5 & 63 04 36 & 18 & 18 & 24 & Rejected: 2nd component flatter, 3rd component extended\\
    J1343+2844 & 13 43 00.2 & 28 44 07 & 6 & 6 & - & Rejected: 2nd component extended, steeper\\
    J1540+1447 & 15 40 49.5 & 14 47 46 & 4 & 5 & 28 & Rejected: 2nd component extended, steeper\\
    J1641+5115 & 16 41 55.7 & 51 15 47 & 9 & 10 & - & Rejected: 2nd component extended, steeper\\
    J2139+1027 & 21 39 42.6 & 10 27 43 & 7 & 7 & 14 & Rejected: 2nd component steeper\\
    J2353+3231 & 23 53 20.9 & 32 31 44 & 14 & 14 & - & Rejected: 2nd component extended, steeper\\
    \hline
  \end{tabular}
\end{table*}

We selected 18 candidates which were fitted by multiple compact components 
in the initial CLASS VLA snapshots, and for which the second brightest 
component was $10-30$ times fainter than the brightest. New observations of 
these sources were made with the VLA in A configuration at 5 and 8.5~GHz on 
2006 February 12 and 2006 February 19. 
Each source was observed for a few minutes in several blocks at different 
hour angles, giving better u-v coverage than the initial snapshots.
Candidates with multiple compact components at 5 and 8.5~GHz
were re-observed at 22~GHz on 2006 March 20 or 2006 April 03.
Most candidates were found to have an extended secondary component,
or differing spectral indices between components, leaving
3 candidates as possible lenses, J0316+4328, J0935+0719 and J0958+2948.
The observations are summarised in Table~\ref{vla}.

J0958+2958 has 2 compact radio sources separated by 8\arcsec, but these
objects are distinct quasars with redshifts of 2.064 and 2.744, ruling out this
system as a gravitational lens \citep{leh01a}. J0935+0719 has a series of radio
observations indicating that it might be a gravitational lens in which the 
source is a compact symmetric object, and is mentioned as the 
candidate J0935+073 in \citet{bro03a}. We find that the flux density 
ratio is 19:1 at 5, 8.5 and 22~GHz, supporting the lensing interpretation. A 
definitive answer on the nature of J0935+0719 requires better optical data, 
both imaging and spectroscopic, and we do not consider this system further 
in this paper. 

This left J0316+4328 as a possible candidate lacking high resolution
radio imaging. After completing the longer VLA observations we
realised that another source, J0856+4935, looked promising in the initial 
CLASS snapshot. On December 08 2006 and December 09 2006 we observed both 
sources at 6.0~GHz with the Multiple Element Radio-Linked Interferometric 
Network (MERLIN), spending 12 hours on each source. J0856+4935 had a diffuse
secondary component and was rejected as a lens, while J0316+4328 showed 2
compact components and remained as a lens candidate with increased confidence.

\begin{figure}
  \includegraphics[width=0.5\textwidth]{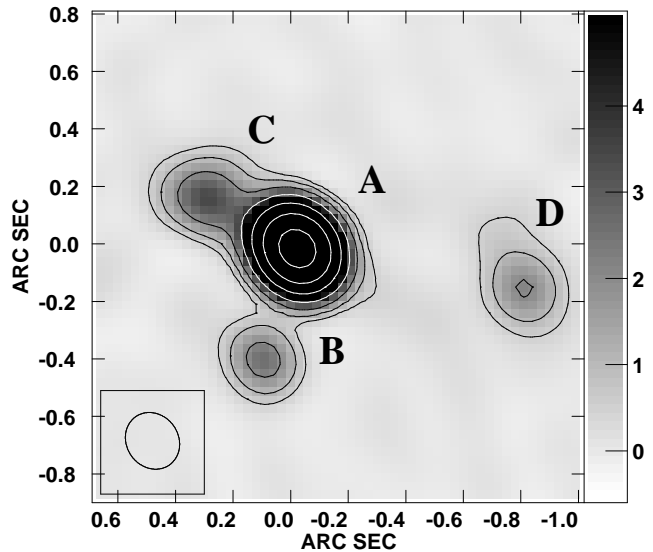}
  \caption{8.5~GHz VLA A array map of J0316+4328, made with super-uniform 
    weighting. Components A and B are probably 2 images of a lensed quasar, 
    with a flux density ratio of $17.0\pm0.6$.
    C and D are probably lobes situated either side of the
    quasar core which lie too far from the lensing galaxy to form multiple 
    images.
    The  off-source rms is 0.09~mJy/beam and the beam is 
    $0\farcs20\times0\farcs18$ at a position angle of $35^\circ$. The 
    co-ordinates are offset from 03~16~50.933~+43~28~19.31, the greyscale is 
    in mJy/beam, and the contours increase in factors of 2 from 0.5~mJy/beam.
    \label{vla8}
  }
\end{figure}

\begin{figure}
  \includegraphics[width=0.5\textwidth]{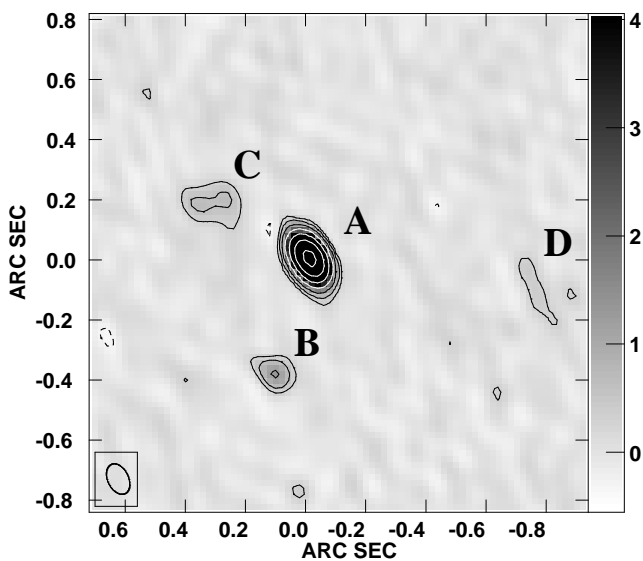}
  \caption{22.5~GHz VLA A array map of J0316+4328.
    Components A and image B are both compact, with
    a flux density ratio of $19.6\pm1.3$.
    With higher resolution, components C and D are clearly extended.
    The off-source rms is 0.09~mJy/beam and the beam is 
    $0\farcs11\times0\farcs07$ at a position angle of $25^\circ$. The 
    co-ordinates are offset from 03~16~50.933~+43~28~19.31, the greyscale is 
    in mJy/beam, and the contours increase in factors of 2 from 0.3~mJy/beam.
    \label{vla22}
  }
\end{figure}

\begin{figure}
  \includegraphics[width=0.5\textwidth]{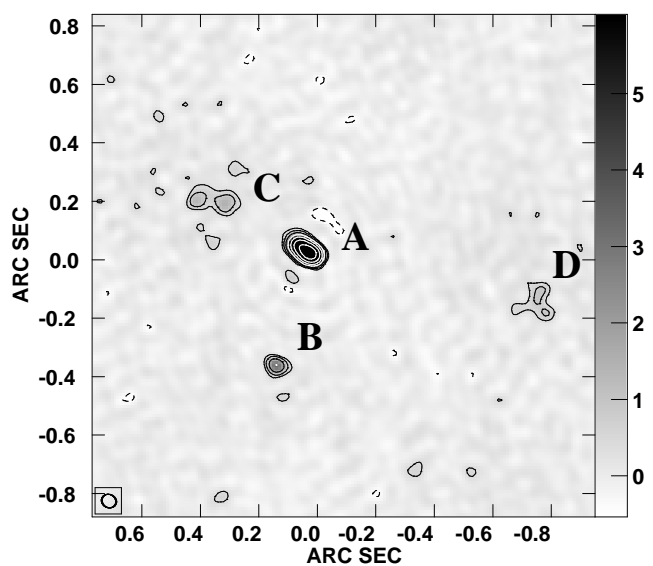}
  \caption{6.0~GHz MERLIN map of J0316+4328.
    Component B is compact, component A is marginally extended 
    (major axis 0\farcs051 at a position angle of $54^\circ$).
    The flux density ratio of components A and B is $19.2\pm0.8$, 
    taking the total flux density of A. 
    Components C and D are resolved by MERLIN. 
    The off-source rms is 0.13~mJy/beam and the beam is 
    $0\farcs051\times0\farcs045$ at a position angle of $61^\circ$. The 
    co-ordinates are offset from 03~16~50.93~+43~28~19.3, the greyscale is 
    in mJy/beam, and the contours increase in factors of 2 from 0.4~mJy/beam.
    \label{merlin6}
  }
\end{figure}

\begin{table*}
  \centering
  \caption{Gaussian fits to the components of J0316+4328,
    giving positions relative to component A in the 22~GHz VLA map, 
    integrated flux densities and sizes (deconvolved major axes). 
    Components C and D were clearly resolved in the 6.0~GHz MERLIN map
    (see Figure~\ref{merlin6}), so in this case 2 gaussians were fitted.
    The flux density is the total flux density of the 2 components, the size 
    is their separation.
    B is unresolved in each map, while A is marginally resolved with MERLIN.
    The off-source rms values were 0.13, 0.09 and 0.09~mJy/beam and the beam 
    sizes were $0\farcs051\times0\farcs045$, $0\farcs20\times0\farcs18$ and
    $0\farcs11\times0\farcs07$ in the 6.0, 8.5 and 22.5 GHz maps, respectively.
    \label{mvfits}
  }
  \begin{tabular}{@{}lrrrrrrrr@{}}
    \hline
    Component & R.A. & Dec. & \multicolumn{2}{r}{6.0 GHz MERLIN} & \multicolumn{2}{r}{8.5 GHz VLA} & \multicolumn{2}{r}{22.5 GHz VLA}\\
    & & & Size & $S$ (mJy) & Size & $S$ (mJy) & Size & $S$ (mJy)\\
    \hline
    A & 0\farcs00 & 0\farcs00 & 0\farcs051 & 61.91 & 0\farcs06 & 45.75 & 0\farcs05 & 26.60\\
    B & 0\farcs11 & -0\farcs38 & 0\farcs017 & 3.23 & 0\farcs05 & 2.69 & 0\farcs05 & 1.36\\
    C & 0\farcs32 & 0\farcs18 & 0\farcs09 & 5.26 & 0\farcs17 & 4.89 & 0\farcs15 & 2.09\\
    D & -0\farcs76 & -0\farcs12 & 0\farcs07 & 4.08 & 0\farcs21 & 3.25 & 0\farcs29 & 1.21\\
    \hline
  \end{tabular}
\end{table*}

The maps of J0316+4328 with the VLA and MERLIN are presented as 
Figures \ref{vla8}, \ref{vla22} and \ref{merlin6}, while fits to the components
are presented in Table~\ref{mvfits}.
We also mapped the system in Stokes Q, U and V, using the MERLIN data, 
and detected no polarised emission at an rms of 0.13~mJy/beam.
J0316+4328 has 2 compact components (A and B) with flux
density ratios of $19.6\pm1.3$, $17.0\pm0.6$, $19.2\pm0.8$ at 
6.0~GHz, 8.5~GHz and 22~GHz, respectively. Errors are 1-$\sigma$ values
taken from the map rms.
The 2 steeper spectrum components (C and D) either side of 
A and B are extended with the VLA and clearly resolved by MERLIN.
These are probably lobes situated either side of the radio source's core which 
lie too far from the lensing galaxy to be multiply imaged.

\begin{figure}
  \includegraphics[width=0.5\textwidth]{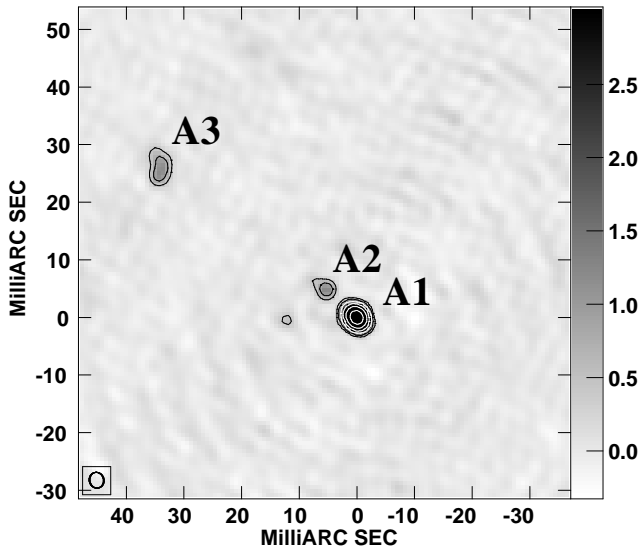}
  \caption{8.4~GHz VLBA map of the A component in J0316+4328, omitting the 
    Mauna Kea, Hancock and St. Croix antennas.
    At VLBI scales component A splits into an unresolved 
    core A1 and jet components A2 and A3 at distances of 7 and 43~mas and 
    position angles of $48^\circ$ and $53^\circ$, respectively. This matches 
    the 51~mas size and $54^\circ$ position angle of component A in the 
    6.0~GHz MERLIN map (Figure~\ref{merlin6}). 
    The off-source rms is 0.08~mJy/beam and the beam is $2.8\times2.5$~mas at 
    a position angle of $16^\circ$. The co-ordinates are offset from component
    A1, the greyscale is in mJy/beam, and the contours increase
    in factors of 2 from 0.4~mJy/beam.
    \label{vlba8a}
  }
\end{figure}

\begin{figure}
  \includegraphics[width=0.5\textwidth]{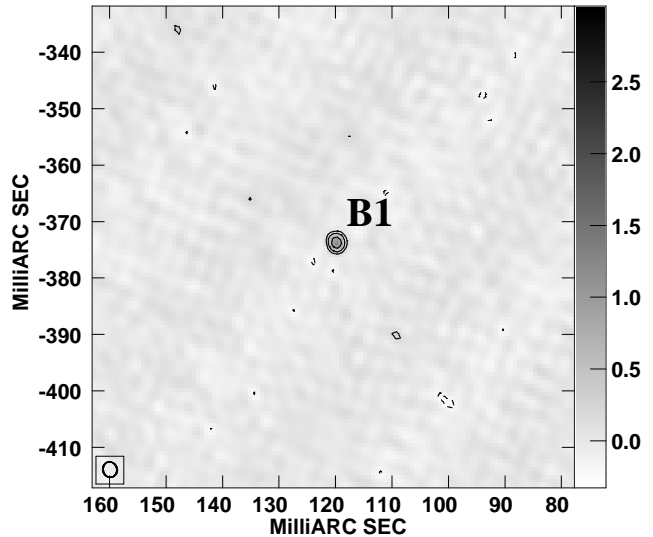}
  \caption{8.4~GHz VLBA map of the B component in J0316+4328, omitting the 
    Mauna Kea, Hancock and St. Croix antennas.
    It appears as a single compact point source B1, with a flux density
    ratio relative to A1 of $23.7\pm1.5$. 
    The off-source rms is 0.065~mJy/beam and the beam is $2.8\times2.5$~mas at 
    a position angle of $16^\circ$. The co-ordinates are offset from the 
    brightest component A1 (shown in Figure~\ref{vlba8a}), the greyscale is in 
    mJy/beam, and the contours increase in factors of 2 from 0.2~mJy/beam.
    \label{vlba8b}
  }
\end{figure}

\begin{table*}
  \centering
  \caption{Gaussian fits to the components of J0316+4328 in the VLBA maps, 
    giving positions relative to A1, integrated flux densities and sizes 
    (deconvolved major axes).
    The blank field rms values were (0.08,0.065)~mJy/beam in the A and B 
    fields, respectively. The beam size was $2.8\times2.5$~mas.
    A1, A2 and B1 are compact.
    \label{vlbafits}
  }
  \begin{tabular}{@{}lrrrr@{}}
    \hline
    & R.A. (mas) & Dec. (mas) & Size (mas) & $S$ (mJy)\\
    \hline
    A1 & 0.0 & 0.0 & 1.3 & 23.7\\
    A2 & 5.4 & 4.9 & 2.2 & 1.6\\
    A3 & 34.1 & 25.9 & 4.5 & 3.0\\
    B1 & 120.4 & -379.9 & 0.0 & 1.0\\
    \hline
  \end{tabular}
\end{table*}

We observed J0316+4328 with the Very Long Baseline Array (VLBA) at 8.4~GHz 
on 2006 July 16. The observation was made at an observing rate of 256~Mb/s 
and included 113 minutes on J0316+4328.
Self-calibration was used to derive the phase solutions, and it was
not possible to fit good solutions on the longest baselines, so the
Mauna Kea, St. Croix and Hancock antennas were omitted. Components A
and B (Figures \ref{vlba8a} and \ref{vlba8b}) were found to have
reduced flux densities compared to those in the VLA maps
(Table~\ref{vlbafits}), presumably due to resolving out extended
emission. No emission was detected at the locations of components C and D;
they were completely resolved out by the VLBA.
Component A has 84\% of its flux density in a central point source A1 
and the remainder in jet components A2 and A3. Component B has a single 
point source sub-component B1, with a flux density ratio relative to A1 of 
$23.7\pm1.5$, matching the 6~GHz and 22~GHz ratios at the 2-$\sigma$ level. 

Some simple lens models which fit the bright images place the central image
only $\sim20$~mas from image B. While these models are severely 
underconstrained (due to the dearth of information on the lens galaxy),
it is plausible that the VLA and MERLIN maps would blend image B with the 
central image, meaning that only the VLBA maps constrain the central image.
No source was seen between images A and B, with a 5-$\sigma$ limit of 0.33~mJy,
although interpreting a non-detection over 
many resolution elements is not straightforward, see \citet{zha07a}.
 
\section[]{Discussion}

\begin{figure}
  \includegraphics[width=0.5\textwidth]{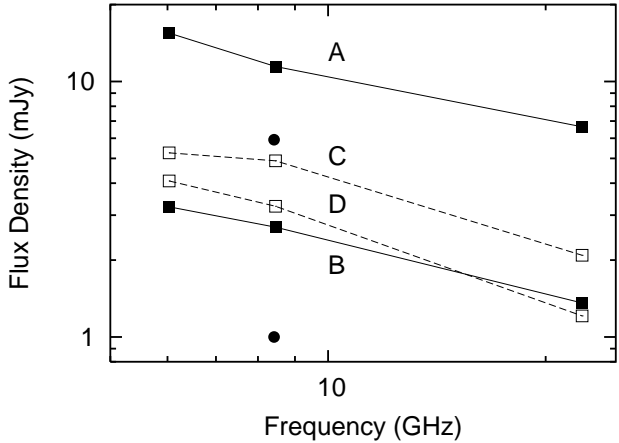}
  \caption{Spectra of the components of J0316+4328 from the VLA and MERLIN
    observations. A and B (solid squares, solid lines) have the same spectrum 
    from 6~GHz to 22~GHz, while C and D (hollow squares, dashed lines) are 
    noticeably steeper. The VLBA flux densities of A1 and B1 are plotted as 
    solid circles. Component A and A1 flux densities are multiplied by 0.25. 
    \label{fluxgraph}
  }
\end{figure}

Although we lack high-resolution optical data, we are very confident that
J0316+4328 is a gravitational lens. Components A and B are very likely to be 
lensed images of the same quasar, as they are compact at resolutions 
from 0\farcs2 to 0\farcs05 and exhibit the same flux density ratio over a 
factor of 3.5 in frequency (Figure~\ref{fluxgraph}). At 3~mas resolution 
substructure is seen only in the brighter, more magnified image.
All but one of the CLASS candidates with multiple compact components in VLBA 
maps were confirmed as lenses \citep{bro03a}. The exception, B0827+525, had
significantly different spectra for the 2 components and is likely
to be a binary quasar \citep{koo00b}.

We plan to obtain a VLBI map with rms $\sim0.02$~mJy, which would 
definitively confirm the lensing interpretation by detecting parity-reversed 
substructure in image B matching that in image A. If B is a lensed image 19 
times fainter than A, it should include sub-components B2 and B3 with flux 
densities of 0.08 and 0.16~mJy, and separations from B1 of 1.7 and 10~mas. 
This map would also improve the flux density measurement of B1 
(which currently limits the precision of the VLBI flux density ratio), 
and might even detect a central image.

We are also pursuing high resolution optical images with the William
Herschel telescope and spectroscopic observations with the Lick telescope. 
The USNO B1 catalog shows an optical object with R=19.1 and B=20.3 at
03 16 50.8693 +43 28 19.880, an offset of -0\farcs69 in right ascension 
and +0\farcs55 in declination from image A in the MERLIN map.
The USNO catalog has astrometric accuracy of 0\farcs20 \citep{mon03a}
and this object has errors of 0\farcs38 in right ascension and 0\farcs10 in 
declination, while the MERLIN map has astrometric accuracy and positional
error both better than 0\farcs05. The optical and radio positions 
are consistent at the 2-$\sigma$ level, but it is not possible to identify 
the USNO object as a lensed image, the lens galaxy, or a blend of both. 
High resolution optical images should confirm the lensing interpretation,
and allow detailed lens modelling.

For isothermal profiles with moderate ellipticity or shear, the lensing 
cross-section for asymmetric doubles (magnification ratio $10-20$) is 
$\sim20\%$ that of quads and symmetric doubles (magnification ratio $1-10$)
combined. With 21 quad or symmetric double lenses in CLASS, 4-5 asymmetric 
doubles would be expected.\ \ 3 such systems have been found, counting
the confirmed lens B1030+074 and the 2 strong candidates J0316+4328 and 
J0935+0719.

Our lens modelling must accommodate the extended, steep-spectrum components C 
and D. Their radio spectra are different from each other as well as from A and 
B (Figure~\ref{fluxgraph}), and their separation is 1\farcs12 as opposed to 
0\farcs40 for images A and B, so C and D are unlikely to be a pair of lens 
images. Their orientation either side of image A, and the fact that 
substructure in A points towards the brighter component C, argues that they 
are diffuse lobes lying either side of the source's core, in a typical double 
radio source configuration. With the lens image separation of 0\farcs40
arguing for a relatively less massive lens galaxy with a small lensing
cross-section, it should be possible to find models where the source's lobes 
lie outside the multiply-imaged region of the source plane, while the core
lies just within it. The configuration actually resembles that of the first 
lens system B0957+561 \citep{wal79a,rob79a}.

Our modelling will give strong constraints on the lens galaxy's density 
profile. For near isothermal models which are a good approximation to 
most lens galaxies \citep{rus03a,koo06a}, images A and B form at radii 
differing by a factor of 15-20 and constrain the density profile over this 
large range of galactocentric radius. The constraints will be particularly 
strong if substructure is detected in each image, as in the case of B1152+199
\citep{rus02a}. Also, central images become brighter as 
image B becomes fainter, relative to image A. Compared to quad or
symmetric double lenses, J0316+4328 is more likely to show a central 
image in sensitive VLBI maps, and even an upper limit on the central image 
flux density will be a more stringent model constraint.

\section[]{Conclusion}

We have detected a probable gravitational lens with 2 bright images at a 
considerably higher flux density ratio than any previously known double lens.
Improved VLBI and high-resolution optical observations are underway. This 
highly asymmetric lens should give unusually stringent 
constraints on the density profile of the lens galaxy.

\section[]{Acknowledgments}

MERLIN is a National Facility of STFC operated by the University of Manchester.
The National Radio Astronomy Observatory is a facility of the National Science 
Foundation operated under cooperative agreement by 
Associated Universities, Inc.
Support for WJS and the VLBA observation of J0316+4328 were provided by the 
Research Experience for Undergraduates program at NRAO.
This work was supported in part by the European Community's Sixth Framework 
Marie Curie Research Training Network Programme, Contract No. 
MRTN-CT-2004-505183 "ANGLES."

\bibliographystyle{mn2e}
\bibliography{boyce_J0316}

\label{lastpage}

\end{document}